\documentclass[%
preprint,
 amsmath,amssymb,
 aps,
pra,
floatfix
]{revtex4-1}
\usepackage{lineno}
\usepackage{graphicx}
\usepackage{dcolumn}
\usepackage{bm}
\usepackage{xcolor}
\usepackage{titlesec}
\titlespacing*{\section}{0pt}{0.1\baselineskip}{0.2\baselineskip}

\begin{document}

\title{Frequency-selective terahertz wave amplification by a time-boundary-engineered Huygens’ metasurface}

 \author{Fu Deng$^1$}
 \thanks{The authors contributed equally to this work}
 \author{Fengjie Zhu$^2$}
 \thanks{The authors contributed equally to this work}
 \author{Xiaoyue Zhou$^1$}
 \author{Yi Chan$^1$}
 \author{Jingbo Wu$^{2,3}$}
 \author{Caihong Zhang$^{2,3}$}
 \author{Biaobing Jin$^{2,3}$}
 \author{Jensen Li$^{1,4}$}
 \author{Kebin Fan$^{2,3}$}
  \email{kebin.fan@nju.edu.cn}
 \author{Jingdi Zhang$^{1,4}$}
 \email{jdzhang@ust.hk}
 \affiliation{
 $^1$Department of Physics, Hong Kong University of Science and Technology, Kowloon, Hong Kong SAR, China \\
 $^2$Research Institute of Superconductor Electronics (RISE) $\&$ Key Laboratory of Optoelectronic Devices and Systems with Extreme Performances of MOE, School of Electronic Science and Engineering, Nanjing University, Nanjing 210023, China \\
 $^3$Purple Mountain Laboratories, Nanjing 211111, China \\
 $^4$William Mong Institute of Nano Science and Technology, Hong Kong University of Science and Technology, Kowloon, Hong Kong, China
 }

\date{\today}

\begin{abstract}
Ultrafast manipulation of optical resonance can establish the time-boundary effect in time-variant media leading to a new degree of freedom for coherent control of electromagnetic waves. Here, we demonstrate that a free-standing all dielectric Huygens’ metasurface of degenerate electric and magnetic resonances can prompt the broadband near-unity transmission in its static state, whereas it enables wave amplification in the presence of time boundary. The time boundary is realized by femtosecond laser excitations that transiently inject free carriers into the constituent meta-atoms for dynamic removal of a pre-established two-fold degeneracy. We observe that the transmittance in the photo-excited Huygens’ metasurface can exceed unity transmittance, i.e., THz wave amplification, by a factor over 20\% in intensity at frequencies tunable by varying the arrival of time boundary with respect to that of the seed terahertz pulse. By numerical simulations and analysis with time-dependent coupled mode theory, we show that the wave amplification results from the ultrafast Q-switching and shift in resonant frequencies. This work demonstrates a new approach to achieve tunable amplification in an optical microcavity by exploiting the concept of time-variant media and the unique electromagnetic properties of Huygens’ metasurface. 
\end{abstract}

\maketitle

\section{Introduction}

The simultaneous interaction of the electric and magnetic components of light in metasurfaces has profoundly expanded the frontier of electromagnetic wave manipulation \cite{chen2018huygens, shamkhi2019transparency, miroshnichenko2015nonradiating, shamkhi2019transverse, decker2015high, fan2018phototunable, fan2022active}. By exquisite control over the geometry, spatial distribution, and orientation in array of resonators that are metallic or dielectric, the pertinent local electric and magnetic multipoles can facilitate tunable operating frequency over a remarkably spectral range. This capability has brought forth a diverse range of intriguing phenomena, including mode-degeneracy-induced transparency \cite{chen2018huygens,shamkhi2019transparency}, anapole effect \cite{miroshnichenko2015nonradiating}, perfect absorber or directional scattering \cite{shamkhi2019transparency, shamkhi2019transverse, landy2008perfect} and exceptional points \cite{moritake2023switchable}. Among all the base materials for constructing metasurface, dielectric metasurfaces outstand as an ideal platform for demonstrating low loss and, meanwhile, dedicated control over various fundamental properties of light, e.g., amplitude, phase \cite{jahani2016all, arbabi2015dielectric}, polarization \cite{arbabi2015dielectric} and propagating direction \cite{shamkhi2019transverse} at an unprecedented level \cite{shamkhi2019transverse, jahani2016all}. Its potential for tailoring electromagnetic wave bears new opportunities for photonic technologies and paves the way for advancements in various areas of application, such as optical communication, modulators, sensing, structured beams and holograms generation upon geometrically varying the constituent resonators\cite{fan2022active}.
  
Besides electromagnetic wave manipulation by careful patterning resonator arrays in the real space, one could leverage ultrafast dynamic response of the underlying base material to intense pulse laser excitation (a near-IR pump beam), which can drastically modify in the time domain \cite{shaltout2019spatiotemporal, galiffi2022photonics13,yin2022floquet14}, the effective polarizability of a metasurface as a novel route for active control over an accompanying light wave (a THz seed beam). Such configuration can establish the time delay of the control and controlled pulses ($t_{pp}$) as the new dimension for regulating the light-metasurface interaction, i.e., time-boundary engineering. When the duration of the excitation-pulse-induced dynamic process is sufficiently short, an abrupt temporal discontinuity in the dielectric response of medium, often referred to as the temporal or time boundary, can be established to provide instantaneous modulation to waveform of the seed pulse, leading to many remarkable phenomena,  such as time-modulation-induced non-reciprocity \cite{yu2009complete15, sounas2017non16, guo2019nonreciprocal17}, frequency tuning and conversion \cite {zhou2020broadband18, moussa2023observation19, bohn2021spatiotemporal20, lee2018linear21, tirole2023double22} and time reversal \cite{bacot2016time23, vezzoli2018optical24}, which are not attainable to conventional resonators of stationary response function. To date, much effort has been made in theoretical prediction \cite{bakunov2021constitutive25, bakunov2021light26, solis2021time27} and experimental demonstration of novel properties relevant to time-boundary engineering \cite{shcherbakov2019photon28, shcherbakov2019time29, lemasters2021deep30, karl2020frequency31, cong2021temporal32, lee2022resonance33}. In particular, the time-boundary-engineered metasurfaces of high-quality factors show nontrivial time-dependent responses in the linear regime \cite{karl2020frequency31, cong2021temporal32, shcherbakov2017ultrafast34}, for instance, the effect of frequency conversion and photon braking, as well as photon acceleration in the nonlinear regime \cite{shcherbakov2019photon28, liu2021photon35, zubyuk2022externally36}. Despite the significant progress made in time-boundary-engineered dielectric metasurfaces, a clear-cut amplification effect at frequencies tunable on ultrafast timescale has not been demonstrated. The reason being is that most studies by far manipulate only a single mode in the metasurface, inevitably introducing energy loss due to its impedance unmatched to the free space. However, in principle, one can exploit the unique Huygens’ metasurface offering degenerate dual-mode interaction to light for a matched impedance to remove the handicap of unnecessary energy loss \cite{chen2018huygens} and facilitate the predicted amplification effect \cite{knorr2022intersubband37}.

Here, in a free-standing Huygens’ metasurface (HM), we unambiguously demonstrate that the terahertz (THz) wave at selected frequencies can be amplified on its traversing the sample and in the presence of a time boundary, established on arrival of a femtosecond intervention by the near-IR pump pulse. Owing to the dual-channel dynamic modification to electromagnetic response by the time boundary, a clear net energy gain, indicating amplification, was achieved in the time-boundary-engineered HMs, as evident by an above-unity transmittance at frequencies tunable by the pump-seed delay; the transmitted wave exceeds the incident free-space THz pulse by more than 10\% in magnitude, equivalently 20\% in power. The amplification effect is manifested in distinct spectral fringes, of which the periodicity can be continuously tuned in response to a varying pump-seed-pulse delay. Such effect comes into existence almost instantaneously and remains over a time window as wide as 35 ps, corresponding to a broad tuning range in the frequency domain. Our experimental findings are in excellent agreement with both full-wave numerical study and analytical study using time-dependent coupled mode theory, showing that the amplification effect originates from light-induced ultrafast shift in resonant frequencies and the simultaneous suppression of the quality factors (Q-factors) of the degenerate electric and magnetic dipole modes. Although demonstrated at THz frequencies, the underlying mechanism presented herein can readily be extended to other frequencies for optical amplification by leveraging optical cavities of degenerate dual-mode resonances. This concept may find promising new applications in metasurfaces as gain media by means of temporally tailored light-matter interactions. 

\section{Time-boundary effect in a Huygens’ metasurface}

\begin{figure}[htbp]
\centering
{\includegraphics[width=12cm]{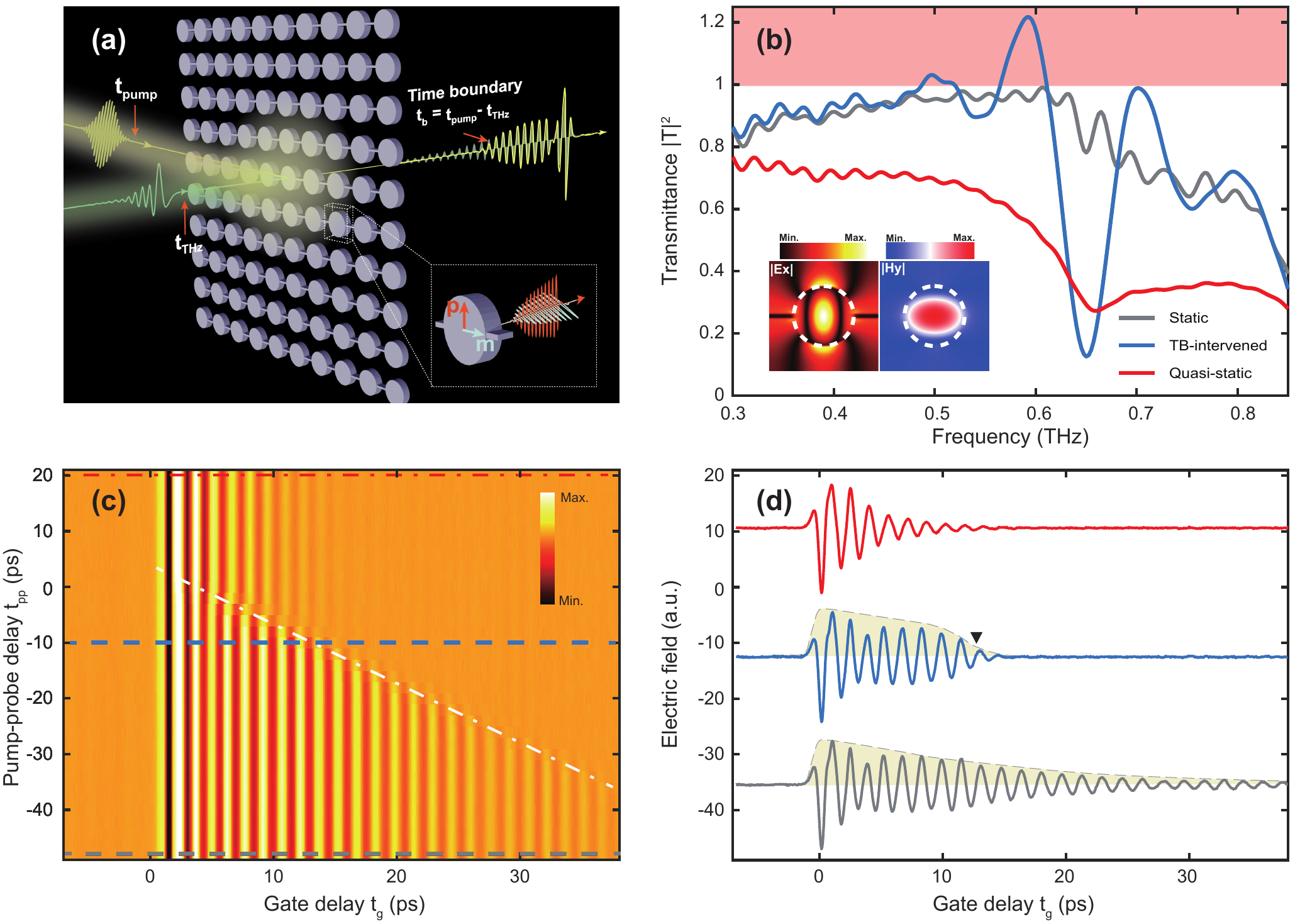}}
\caption{Amplified transmission from time-variant Huygens’ metasurface. (a) Schematic illustration of a free-space THz field (green) passing through a time-variant Huygens’ metasurface and the resultant transmitted pulses without (green shade) and with (yellow line) the presence of time boundary. The time boundary is imposed by excitation from an optical pump pulse (left, yellow pulse). The close-up of individual resonator shows time-boundary-modulated local dipole moments (red, electric mode; white, magnetic mode). (b) Frequency-domain transmittance spectra of the Huygens’ metasurface in its static state (gray), time-boundary-intervened state at negative delay of -10 ps (blue) and the quasi-static state at positive delay (red). The shaded area indicates the transmittance larger than unity ($|T|^2>1$). The insets are the in-plane electric (E$_x$) and magnetic (H$_y$) field distributions of the HM unit cell at the degenerate resonant frequency (\textasciitilde 0.65 THz). (c) Two-dimensional time-domain measurement on THz waveforms transmitted through the sample as a function of gate ($t_g$) and pump-probe delay ($t_{pp}$), at an excitation fluence of 20 $\mu$J/cm$^2$. The white dashed line indicates the moving time boundary. Horizontal line cuts indicate time delays in order of colors of spectra shown in (b). (d) Far-field time-domain signals of the THz pulse after interaction with HM sample in three representative states above (gray, static state; blue, time-boundary intervened state; red, quasi-static state). The shade with dashed outline show the a double-edge signal envelope established jointly by the Lorentzian response and time boundary (black triangle) and a positive-edge signal envelope by Lorentzian response alone.}
\label{Fig1}
\end{figure}

The schematic in Fig. \ref{Fig1}a conceptually shows the time boundary effect on a broadband THz pulse, mediated by an all-dielectric HM. The HM in the equilibrium state displays unity transmittance, due to the impedance match established by the degenerate electric dipole (ED, red arrow) and magnetic dipole (MD, white arrow) modes, as demonstrated in previous works \cite{decker2015high, fan2018phototunable}. In the equilibrium (static) state without photoexcitation, the broadband single-cycle THz seed pulse interacts in two distinct ways depending on the resonance condition. Off-resonance, the matched impedance allows the incident light to freely propagate through the sample at near unity transmission. On-resonance, the incident light can be fully permissible into the dielectric cavity, contained inside the disk and gradually re-radiated in the form of a long-lived oscillatory signal in transmission, in stark contrast to the direct passage of off-resonance spectral component of the THz pulse. Benefiting from the all dielectric configuration, the dissipative loss is negligibly small during the two-step process of trapping and re-radiation of energy. As such, the dedicated HM assures, in static state, near unity transmission in magnitude at all frequencies in the spectral range concerned. Despite their negligible impact on the magnitude spectrum, the excitation of bimodal resonators collectively coerces a total phase shift of 2$\pi$ across the resonance frequency, resulting from even contribution of Lorentzian response by the degenerate electric and magnetic dipoles (Fig. S2 and S4). Furthermore, assured by causality-respecting Lorentz oscillator and coherent broadband interaction at THz regime, we note that a rising-edge intensity profile can be naturally imprinted onto the local in-cavity field, implicitly constituting half a square-wave intensity envelope, essential to amplification effect described below.

To break the static state and create a time boundary, an above-bandgap laser excitation of femtosecond duration was used to bring about inter-band transitions and a surging free carrier density in the all-silicon-based metasurface. The optical excitation instantaneously lifts the degeneracy in ED and MD modes and compromises the Q-factor. Such ultrafast transition from HM and non-HM type sample turns the electric and magnetic response functions $\varepsilon(\omega)$, $\mu(\omega)$ into strongly time-dependent quantity \cite{morgenthaler1958velocity38} at THz frequencies, which effectively quenches the resonating in-cavity field and results in a falling-edge in the re-radiated free-space field (Fig. \ref{Fig1}a). Alongside the inherent rising-edge modulation pre-established in the HM before entering the non-equilibrium state, the time boundary completes a square-wave modulation to the local sinusoidal fields, borne bimodally by the resonating electric and magnetic dipoles. In the presence of such modulation, energy contained at the degenerate resonance frequency can effectively be transferred to off-resonance frequencies, precisely determined by temporal width of the square-wave envelope (Figure \ref{Fig1}d). It is essentially a single-slit time diffraction, as the counterpart of recently demonstrated time-domain double-slit interference \cite{tirole2023double22}.

Particulars of experimental implementation are as follows. The free-standing HM used in this study are made from a silicon disk array with a diameter (D) of 181 $\mu$m, thickness (H) of 80 $\mu$m, and period (P) of 330 $\mu$m (see Supplementary 1 for details). All silicon disks are made completely free-standing and connected by bridging fins with a width of 12.3 $\mu$m (see Fig. \ref{Fig2}b for scanning electron microscope image). The substrate-free configuration made possible the experimental confirmation on absolute unity transmittance in static state and prevented the otherwise confounding effect, e.g., enhanced transparency due to the bleaching effect. In a THz pump-probe experiment, an ultrafast optical pump pulse (center wavelength 800 nm, 1.55 eV, duration 35 fs) is employed to set the tunable time boundary, and the broadband THz probe pulse interacts with HM sample for preparing a resonating state first to trap energy inside the bimodal cavity and then to release in the presence of the time boundary. The abrupt modification to the resonating features of the HM by the optically injected carriers is almost instantaneous and remains in force on timescale orders-of-magnitude longer than all other timescales concerned in this work, i.e., width of time boundary, decay time of in-cavity field of ED and MD modes, etc. It is in this way that the pump pulse effectively imposes a step-like time boundary on the sample by driving the sample into a “quasi-static” electronic state that is out-of-equilibrium but stable. The delay between the pump pulse and the THz seed pulse can be continuously varied at a precision of femtosecond, so that the time boundary can rise at the desired instant to intervene the local field oscillation, prompting a falling edge to join the rising-edge-modulated sinusoidal dipole oscillation in the silicon cylinder. After the complex two-beam interaction the HM, the far-field signal in forward direction comprises a square-wave modulated re-radiation at resonance and all other field components directly transmitted at off-resonance frequencies. Its real-time waveform is measured with the standard time-domain THz spectroscopy by electro-optic (EO) sampling, which is critical for visualizing the time boundary and complete extraction of complex optical parameters (see Supplementary 2 for more details). 

An exemplary two-dimensional time-domain measurement is in Fig. 1c showing the evolution of transmitted THz pulse in the presence of a moving time boundary, set by excitation pulse at various pump-probe delays (see Supplementary 2 for more details). Specifically, at negative pump-probe delays (\textcolor{red}{$t_{pp}=-t_b<0$}), the incident THz pulse first launches a high-Q oscillation in the local resonators, which can be detected in the far-field re-radiation featuring a long-lived ringing signal that lasts more than 30 ps (Fig. \ref{Fig1}d). However, on the arrival of the intervening pump pulse, the ringing tail of the THz signal, i.e., the forward-scattering field jointly contributed by the ED and MD modes, becomes fully suppressed upon the emergence of the time boundary, marked by the white dashed line in Fig. \ref{Fig1}c. It is worth noting that time-dependent transmission spectra in both amplitude and phase can be simultaneously obtained by referencing the complex spectra of a free-space THz pulse with that of a pulse traversed HM, both acquired by direct Fourier transformation of their respective time-domain signals (see Supplementary 2 for details). The complete measurement on the complex spectra at THz frequencies enables an exhaustive optical parameter extraction, which is otherwise challenging for time-boundary experiments in the visible and near-IR spectral range\cite{shcherbakov2019time29, lemasters2021deep30, karl2020frequency31}. Furthermore, largely due to the propitious disparity in durations of THz pulse (picosecond) and the excitation pulse (femtosecond), ultrafast spectroscopy at THz frequencies outstands as the ideal tool for making definitive coherent control and detailed investigation of the time boundary effect in HM detailed as follows.

\section{Time-boundary enabled tunable terahertz amplification}
To highlight the intriguing aspect of the time-boundary-engineered HM, representative spectra with and without the time boundary are selected from the two-dimensional optical-pump-THz-probe measurements (Fig. \ref{Fig1}b). A comparison of magnitude spectra between static and excited states can readily feature the amplification of the THz wave at frequencies tunable by the pump-probe delay. In the absence of time boundary, the HM is in static state and displays a near unity transmittance in proximity to the resonance frequency at 0.65 THz, for the impedance is matched by introducing the ED-MD degeneracy in the lossless silicon disk. The degeneracy is also confirmed by numerical simulation, showing that the ED and MD modes perfectly overlap in resonant frequency but create electromagnetically distinct local field distributions at the unit cell level (Fig. \ref{Fig1}b insets). In the presence of time boundary ($t_{pp}<0$), we observe the transmittance exceeding unity ($|T|^2>1$), indicating amplification of the THz field on traversing the HM. The maximum amplification rate achieved is approximately 20\% in power with respect to that of the incident free-space THz wave. We stress that the sample used in our work is a free-standing HM sample, thus the larger-than-unity signal means the incident THz pulse gains energy at selective frequency, in contrast to prior experiments \cite{karl2020frequency31, cong2021temporal32} that employ substrate-supported metasurface samples inherently inflicted by the reflection loss. However, at positive delays ($t_{pp}>0$), i.e., pump pulse arrives earlier than the THz pulse, the time-boundary effect is clearly vanishing. The sample first enters the photo-doped quasi-static state featuring a substantially larger damping rate than in the true static state, and then interacts with the THz pulse, that broadens the resonance and reduces globally the transmittance. From the observations above, the amplification effect emerges on the intervention condition, that is, the optical switching of HM by the pump pulse should take place within the coherence time of the local field oscillation, which is initiated by the THz probe pulse in advance. 

\begin{figure}[ht!]
\centering
{\includegraphics[width=\linewidth]{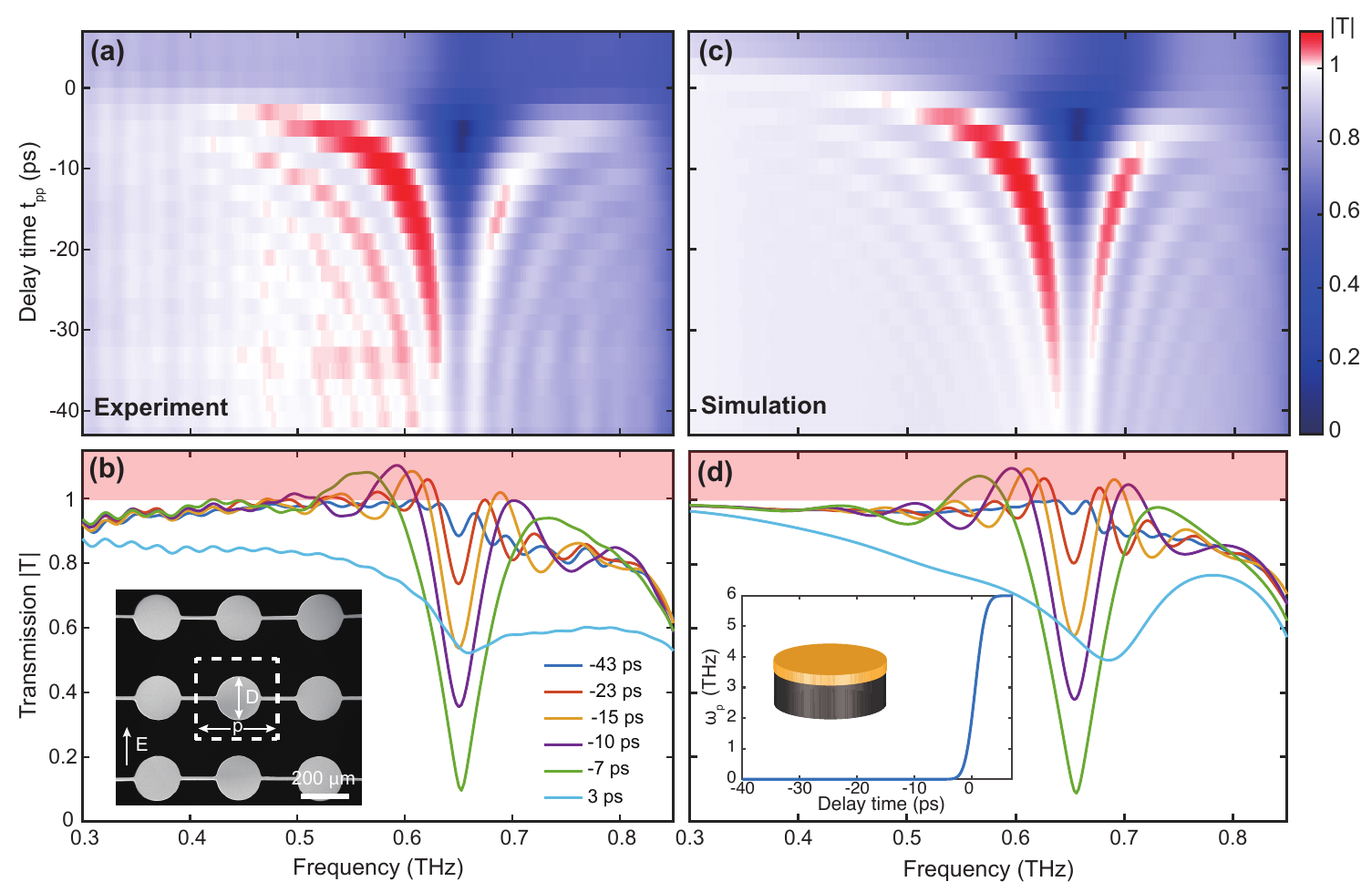}}
\caption{Evolution of the time-variant Huygens’ metasurface, experiment and simulation. (a) Experimental results on time-dependent transmission spectra of Huygens’ metasurface excited at pump fluence of 20 $\mu$J/cm$^2$. The negative delay time means a pump pulse trailing the probe. The color scale is chosen to emphasize the above-unity transmission in red. (b) Transmission spectra at selected pump-probe delays. The inset shows the scanning electron microscope (SEM) image of the free-standing Huygens’ metasurface disk array, 181 $\mu$m in diameter (D), 330 $\mu$m in pitch (P) and 80 $\mu$m in thickness (H). Each row of disks are bridged by a thin rigid supporting line, oriented perpendicular to polarization of incident THz pulse. (c) Simulation results on transmission spectra of Huygens’ metasurface using a time-dependent Drude model. In the quasi-static state, plasma frequency ($\omega_p$) of 6 THz and damping rate ($\gamma_p$) of 1.6 THz are used for the photo-active layer in the silicon meta-element. (d) Simulated transmission spectra at selected pump-probe delays same as in (b). The inset shows the time-dependent plasma frequency of the photo-active silicon layer.}
\label{Fig2}
\end{figure}

For a more comprehensive understanding of the amplification effect developed from the time-boundary-engineered HM, we show in Fig. \ref{Fig2}a the absolute magnitude spectra of the transmission coefficient, i.e., spectra of the pulse through the sample referenced to that of the free-space THz pulse, at different time delays. This figure offers great details into wave amplification when a continuously varying time boundary is at work. In the time-dependent transmission spectra, a fringe pattern emerges in the presence of the time boundary, resulting from the interruption of the pump pulse to the coherent high-Q dipole oscillation launched at an earlier time by the THz seed pulse. It is a prominent feature on two accounts: \textit{transmission peak that exceeds unity and operating frequency that is dynamically tunable}. The first fringe, in proximity to the resonance, surpasses unity ($|T|>1$) on the low frequency side. At still higher frequencies ($>$ 0.8 THz), the high-order mode dominates and compromises the matched impedance, resulting in unwanted energy loss but having little impact on the amplification effect at the lower frequency side. 

In all data sets, the time zero is calibrated and defined by the rising edge of photo-doped carrier dynamics in a separate piece of un-patterned silicon, and the negative delay time corresponds to a THz pulse arriving earlier than the optical pump pulse (see Supplementary 3 for more details). The dynamic quench of high-quality factor of the HM by optical excitation has enabled the persistent presence of well-defined spectral fringes, i.e., wave amplification, starting at a pump-probe delay ($t_{pp}$) as early as -35 ps. The fringe feature then becomes less dense and grows stronger at shorter pump-probe delays. At zero delay, a complete shutdown of the time boundary effect is reached. Meanwhile, the Q-factor of the two resonances diminish rapidly, which can be attributed to the substantial increase in loss that scales with photo-doped carrier density.  This dynamic change in Q-factor is evident from the increasingly strong resonant dip in the line spectra near $t_{pp}$ = 0 ps, signifying an “impaired” Huygens’ metasurface. 

To gain insights, we conducted full-wave numerical simulations to repeat the time-dependent transmission spectra. The meta-element is modeled to be 80 $\mu$m thick free-standing silicon cylinder as a whole. Each cylinder cavity comprises a thick layer of photo-inactive bottom (70 $\mu$m thick) and a thin layer of photo-active top (10 $\mu$m thick) \cite{fan2018phototunable, al2013effect39}, of which thickness is determined by the penetration depth of 1.55 eV excitation photons (inset of Fig. \ref{Fig2}d). In our simulation, the free-carrier response at THz frequencies follows the simple Drude model, defined primarily by the carrier density and scattering rate; both are varying parameters when altering the pump-probe delay or the excitation fluence. At a fixed fluence, the ultrafast dynamics of carrier density ($N_e$), essential for the simulation, can be precisely determined by a separate ultrafast transmission experiment on a thick un-patterned silicon substrate (see Supplementary 3 for more details). The associated plasma frequency ($\omega_p$) subsequently becomes time- and fluence-dependent and follows $\omega_p (t_{pp} )=\sqrt{\frac{N_e(t_{pp}) e^2}{m^*\varepsilon_0}}$, where $t_{pp}$, $N_e$, $e$, $m^*$, $\varepsilon_0$ denote the pump-probe delay, free-carrier density, elementary charge, effective mass and vacuum permittivity, respectively. For instance, at an excitation fluence of 20 $\mu$J/cm$^2$, the plasma frequency $\omega_p$ in our simulation will rise from 0 to 6 THz within 2 ps in the photo-active layer. We incorporate these realistic parameters into the numerical model and simulate time-dependent transmission spectra, shown in Fig. \ref{Fig2}c and d. In a quantitative way, simulation precisely repeats the THz response of the photoexcited all-silicon HM observed by experiments in many aspects, such as the time-boundary-driven dynamics in (i) off-resonance spectral fringes, (ii) above-unity transmission and (iii) on-resonance energy loss feature. With the help of numerical simulation, insights into microscopic dynamics, that are otherwise inaccessible, can be obtained. One finds that the observed amplification effect from the HM roots in an ultrafast suppression of Q-factor and simultaneous removal of the ED-MD degeneracy. It is such microscopic modulation that results in a rapid change in the effective response function of the sample and, in turn, establishes the time boundary, as detailed in the next section. 

To fully explore the effect of the time-boundary-engineered HM in amplifying THz waves, various excitation fluences are used for direct optical control over the carrier density in the meta-elements. The associated 2D transmission spectra are shown in Fig. 3a-d as a function of the frequency and the pump-probe delay, with excitation fluences ranging from 0.25 $\mu$J/cm$^2$ to 5 $\mu$J/cm$^2$. It is evident that, at negative pump-probe delays, the HM excited at the mildest fluences even gives rise to transmission fringes and eventuates in the above-unity transmission at fluences exceeding 1.25 $\mu$J/cm$^2$. Figure 3e shows the line spectra for difference fluences at the pump-probe delay for the largest enhancement in transmission ($t_{pp}$ = -10 ps), which increases with pump fluence from unity to a value of 110\%. In addition, the lossless impedance-matched resonance at 0.65 THz in the static-state transforms into a lossy resonance as the pump fluence increases, implying drastic modification to the Q-factor of the resonating HM due to photo-doped carriers. By replotting the fluence dependence of peak transmission amplitude at a fixed delay, we notice that the maximum amplification rate first increases rapidly at fluences below 1.25 $\mu$J/cm$^2$, then slows down in the range of 1.25-5 $\mu$J/cm$^2$ and finally saturates to 11\% at fluences higher than 5 $\mu$J/cm$^2$. This saturation effect arises primarily from the limit on carrier density amenable to photo-excitation, which poses an upper limit to time boundary effect. 

\section{Theoretical analysis of time-boundary-engineered HM}

For a better understanding of experimental observations, one should acquire instructive information on the ultrafast dynamics in terms of resonant frequencies and Q-factors of the ED and MD modes. Although not experimentally available, it can be obtained by eigenmode analysis after precisely assigning complex eigenvalues to individual meta-elements possessing a photo-active cap layer that is dynamically tunable (see Supplementary 5 for more details). The eigenmode simulation shows, in Fig. \ref{Fig4}a, that the degeneracy is indeed lifted once carrier density in the photo-active layer becomes finite. Upon photoexcitation, the Q-factors of both ED and MD modes quickly fall off as the carrier density approaches 0.2×10$^{17}$ cm$^{-3}$ from below, and become fully suppressed thereafter. Meanwhile, the once-degenerate resonant frequencies of ED and MD modes continuously blue shifts with an increasing carrier density, resulting from a decreasing real component of the permittivity of silicon. The maximum shift in resonant frequencies for both modes are defined as $\Delta\omega_i=\omega_i(N)-\omega_i(0)$, with $i=1$ standing for ED mode and $i=2$ for MD mode. The ED and MD frequency shift are found by the simulation to be 0.05 and 0.2 THz, respectively; that is 7\% and 30\% relative change for ED and MD modes if normalized to their respective resonant frequency in the equilibrium state. To show the relationship between the photo-induced effect on the eigenmodes and the amplification rate, we further performed numerical simulations on the carrier-density dependence of transmission spectra at a fixed delay ($t_{pp}$ = -10 ps). Results in Fig. 4b show that the amplification ($|T|>1$) occurs at a critical carrier density of 0.2$\times$10$^{17}$cm$^{-3}$ and keeps increasing in magnitude at higher carrier densities. The trend described above suggests that the THz wave amplification is realized through a joint effect of the shifting resonant frequency and ultrafast Q-factor switching, to be analyzed using the theory below.  

Incorporating results from eigenmode analysis above into the temporal coupled mode theory (TCMT) \cite{fan2003temporal40}, one can learn more details about the frequency-selective amplification facilitated by the time-boundary effect. It is based on a model involving two nearly orthogonal resonances in an optical cavity, i.e., the ED and MD modes (see Supplementary 5 for more details). The associated temporal coupled mode equations can be described as 
\begin{equation}
\frac{d}{dt}\left(\begin{matrix}a_1\\a_2\\\end{matrix}\right)=-i\left(\begin{matrix}\omega_1^\ast&0\\0&\omega_2^\ast\\\end{matrix}\right)\left(\begin{matrix}a_1\\a_2\\\end{matrix}\right)-i\left(\begin{matrix}\Delta\omega_1^\ast&0\\0&\Delta\omega_2^\ast\\\end{matrix}\right)\left(\begin{matrix}a_1\\a_2\\\end{matrix}\right)f\left(t\right)+K^Ts^+
\end{equation}
where $a_i\ (i=1,\ 2)$ is the effective complex oscillation amplitude of the ED and MD modes respectively; $\omega_i^\ast=\omega_i-i\gamma_i (i=1,\ 2)$ is the static-state ED and MD eigenfrequencies, with $\omega_i$ being the resonant frequencies  and $\gamma_i$ the rate of radiation loss; $\Delta\omega_i^\ast=\Delta\omega_i-i{\Delta\delta}_i$ describes the dynamic shift in resonant frequencies $\Delta\omega_i$ and the change in material loss ${\Delta\delta}_i$; $K=\left[\begin{matrix}i\sqrt{\gamma_1}&\sqrt{\gamma_2}\\-\sqrt{\gamma_1}&-i\sqrt{\gamma_2}\\\end{matrix}\right]$ is the near-field interaction matrix allowing the input field $s^+$ to drive the eigenmodes. In general, the appearance of photo-doped carriers on the top surface of cylindrical microcavity would break the axial inversion symmetry in the static state, forcing the electric and magnetic modes to couple and consequently giving rise to finite-valued off-diagonal components. Under the special circumstance that only a relatively low optical conductivity in silicon is introduced by photo-excitation, the off-diagonal terms in the matrix of differential eigenfrequencies are negligibly small and we, therefore, set them to zeros for simplicity. We introduce a step-like function $f(t)$, i.e., Gauss error function, (see Supplementary 3 for more details) for quantitative description of photo-induced dynamics in key parameters required by the TCMT model. For instance, the time-dependent carrier density in completing the transition from the static to the quasi-static state follows $N(t)=N_{max} f(t)$, where $N_{max}$ is photo-doped carrier density in the quasi-static state at a given excitation fluence. The dynamic changes in resonant frequency and material loss are assumed to scale linearly with the change in carrier density. Both can be readily obtained from the eigenmode analysis at a specified carrier density (Supplementary 5). Combining the static and photo-induced components, real and imaginary parts of the overall dynamic eigenfrequencies follows $\omega_i\left(t\right)=\omega_i+\Delta\omega_if(t)$ and $\delta_i\left(t\right)=\delta_i+{\Delta\delta}_if(t)$, respectively. At the maximum doping level ($N$$\sim$0.56$\times{10}^{17}cm^{-3}$), the best fit to experiment requires the shift in resonant frequencies to be 0.05 THz for $\Delta\omega_1$  (ED) and 0.15 THz for $\Delta\omega_2$ (MD), and the damping rates to vary by 0.026 THz for $\Delta\gamma_1$ and 0.077 THz for $\Delta\gamma_2$. It is noteworthy that the respective dynamic percentage change in resonant frequencies of ED and MD modes ($\Delta\omega_1/\omega_1$; $\Delta\omega_2/\omega_{2}$) are 7.6\% and 23\% in the terahertz Huygens’ metasurface, a figure of merit outdistances its counterparts operating at higher frequencies, e.g., infrared and visible frequencies \cite{shcherbakov2019photon28, karl2020frequency31}. 

Figure \ref{Fig4}c shows the calculated 2D transmission spectra as a function of the pump-probe delay. Its excellent agreement with the experiment on the low frequency side is better resolved in Figure \ref{Fig4}d showing the temporal trace at the frequency of maximum amplification (0.6 THz, green dashed line in Figure \ref{Fig4}c). The deviation from experiments on the high frequency side results from that TCMT-based model herein counts in only two primary modes, but neglects high-order modes and the mode-coupling. Nevertheless, in reproducing the 2D spectra, the TCMT calculation can reveal in great detail the dynamics of ED and MD resonance frequencies (black and orange dashed lines in Fig. \ref{Fig4}c), not easily accessible by other means. Within the picoseconds window for carrier proliferation, the MD mode quickly shifts to a considerably higher frequency than for the ED mode. Concomitantly, the time boundary brings about rapid modulation to rate of material loss ($\delta_i$), i.e., instantaneous attenuation in the Q-factor, enabling in the HM an effective inverse Q-switching process that releases the resonantly contained energy to other frequencies. Furthermore, the TCMT allows for an analytical study of the individual contribution from the ED and MD modes; that is, one can have a clear view of the sole effect of dynamic modulation to the frequency shift and material loss of the ED mode ($\Delta\omega_1$, $\delta_1$) by setting those for the MD mode to zeros, and vice versa. We found the amplification effect depends considerably stronger on ED mode than on MD mode in terms of modulation amplitude to the corresponding parameters ($\Delta\omega$, $\delta$). Meanwhile, the calculation results (Figure \ref{Fig4}d and Supplementary Fig. S8) indicate that the amplification is an effect synergistically determined by all the parameters of the ED and MD modes. The attainable amplification rate is distinctly above that from simply summing contributions by individual modes in the far field and neglecting the bimodal constructive interference. 

Finally, to decisively verify the indispensable role of ED-MD degeneracy in enabling the amplification effect and the presence of propitious interference at the far-field between radiating components of ED and MD origin, two-dimensional transmission measurements of the same kind have been performed but on a regular metasurface without degeneracy. The degeneracy is lifted by expanding the meta-atom diameter (D = 220 $\mu$m), differing from that in a Huygens’ metasurface (see Supplementary 6). The change in geometry, in turn, shifts the ED resonance to lower frequency and makes the MD resonance remain around 0.65 THz. In such a regular metasurface, no above-unity transmission has been observed, as expected.

\section{Discussion}

Except for the above quantitative analysis on the amplification effect, one could understand it from a more intuitive perspective as the following. As described previously, the coordinated double-pulse interaction with the sample–one by the THz pulse and the other by the excitation pulse–establishes in-resonator bimodal moments of electric and magnetic characteristic, that follow in the time domain a square-wave-modulated sinusoidal wave $[\theta(t)-\theta(t-t_b)]cos(\omega_0 t)$, with $\theta(t)$ being the Heaviside step function, $t_b$ the arrival of time boundary and $\omega_0$ the carrier-wave frequency defined at the degenerate resonant frequency. The resultant radiation will feature, in the frequency domain, a spreading of energy to frequencies, centered around but other than the resonance $\omega_0$. It is essentially a temporal single-slit diffraction, with the square-wave rising edge provided inherently by broadband Lorentzian response of the resonator and the falling edge by the time boundary. The unique time-boundary modulation to local field oscillator can, therefore, direct the energy contained by resonating dipoles to be released to other frequencies, shifted from $\omega_0$ by roughly the integer multiples of a characteristic frequency, which scales inversely with width of the square wave ($t_b$). It should also be noted that the falling edge has a finite width of 2 ps in this work, limited by the free-carrier dynamics in the bulk silicon. In principle, it may be further reduced to a sub-picosecond level for a boost in amplification rate if one replaces silicon with semiconductors of direct band gap, \textit{e.g.}, Ge \cite{lemasters2021deep30} or GaAs \cite{lee2022resonance33, shcherbakov2017ultrafast34}., according to our numerical simulation.

To summarize, we have experimentally demonstrated wave amplification in a Huygens’ metasurface by implementing a time boundary through ultrafast optical modulation to the constituent medium. In this proof-of-principle work conducted at THz frequencies, the transmitted power is found to exceed the unity by over 20\% at frequencies tunable by fine control over the delay of time boundary. The amplification effect can persist within a broad time window of several tens of picoseconds, due to the high Q-factor in the static state. Insights gained from the TCMT model and eigenmode analysis further show that the amplification results from the effect of ultrafast shift in natural frequencies and inverse Q-switching mechanism in the HM. These results show the viability of engineering the sub-wavelength scale electromagnetic cavity for light amplification. Although demonstrated in the long-wavelength limit, the mechanism can be extended to optical frequencies and may be applied to develop new light sources, modulators and optoelectronics. Furthermore, similar time-boundary effect can be conjectured to usher in new exciting effects by ultrafast manipulation of complex materials systems, such as polaritonic materials \cite{knorr2022intersubband37, rajabali2021polaritonic41, dunkelberger2018active42, basov2016polaritons43}, high $Q$ metasurface \cite{fan2019dynamic44}, and quantum materials \cite{basov2011electrodynamics45, zhang2014dynamics46}.

\section*{Funding} National Key Research and Development Program of China (Grants No. 2020YFA0309603, 2022YFA1204303); National Natural Science Foundation of China Excellent Young Scientist Scheme (Grant No. 12122416); Hong Kong Research Grants Council (Project No. ECS26302219, GRF16303721); National Natural Science Foundation of China (No.  62275118).

\section*{Acknowledgments} The authors would like to acknowledge Prof. Sir. John B. Pendry for valuable discussion. K. F. acknowledges support from the Fundamental Research Funds for the Central Universities and the Research fund for Jiangsu Key Laboratory of Advanced Techniques for Manipulating Electromagnetic Waves.



\begin{figure}[ht!]
\centering
{\includegraphics[width=\linewidth]{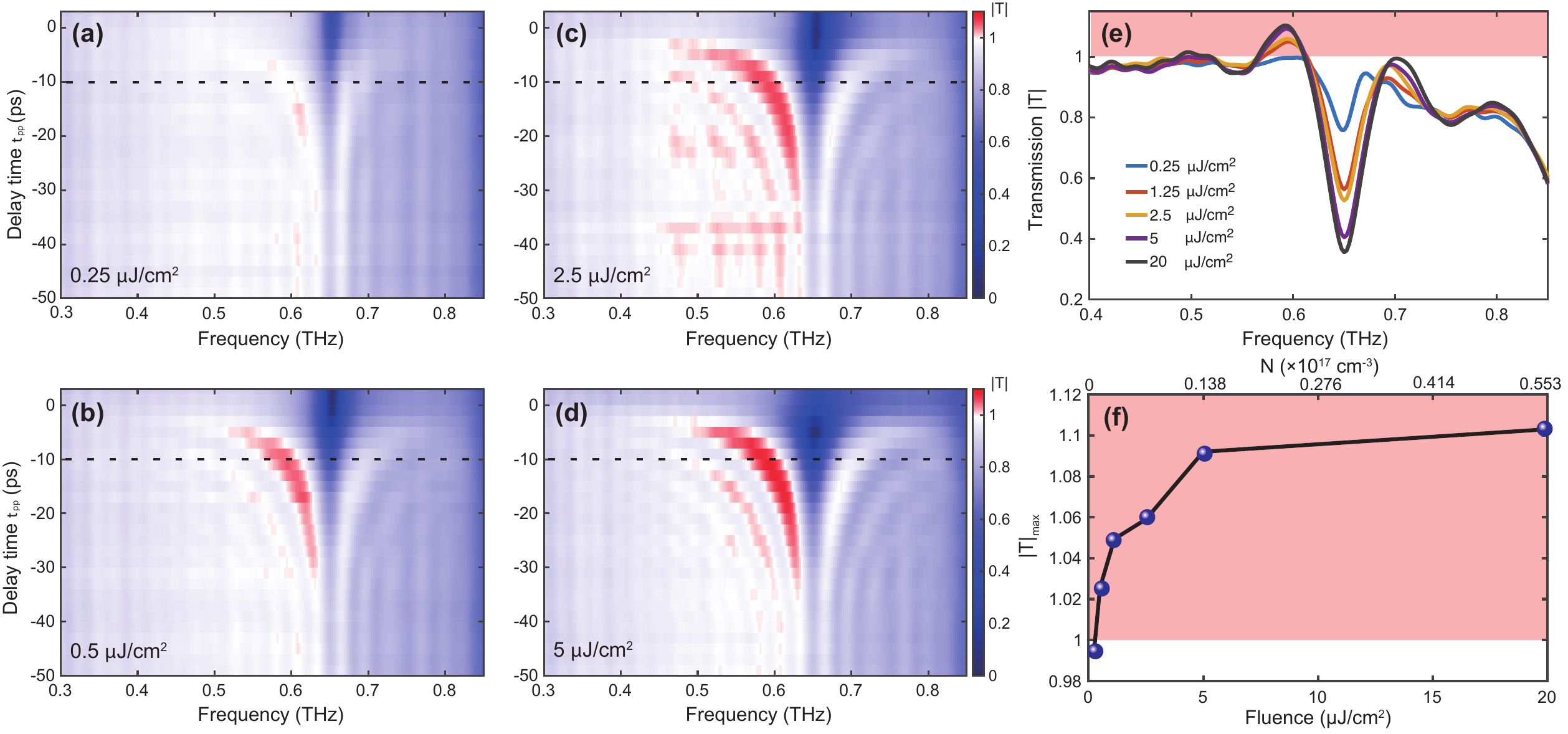}}
\caption{Fluence-dependent two-dimensional transmission spectra of time-variant Huygens’ metasurface. (a)-(d) 2D transmission spectra of Huygens’ metasurface excited by pump beam with an ascending fluence in order of 0.25 $\mu$J/cm$^2$, 0.5 $\mu$J/cm$^2$, 2.5 $\mu$J/cm$^2$ and 5 $\mu$J/cm$^2$. (e) Fluence-dependent transmission spectra at a fixed pump-probe delay of -10 ps. (f) Maximum transmission at various excitation fluence at fixed delay of -10 ps. The shaded area indicates region of amplification. }
\label{Fig3}
\end{figure}

\begin{figure}[ht!]
\centering
{\includegraphics[width=\linewidth]{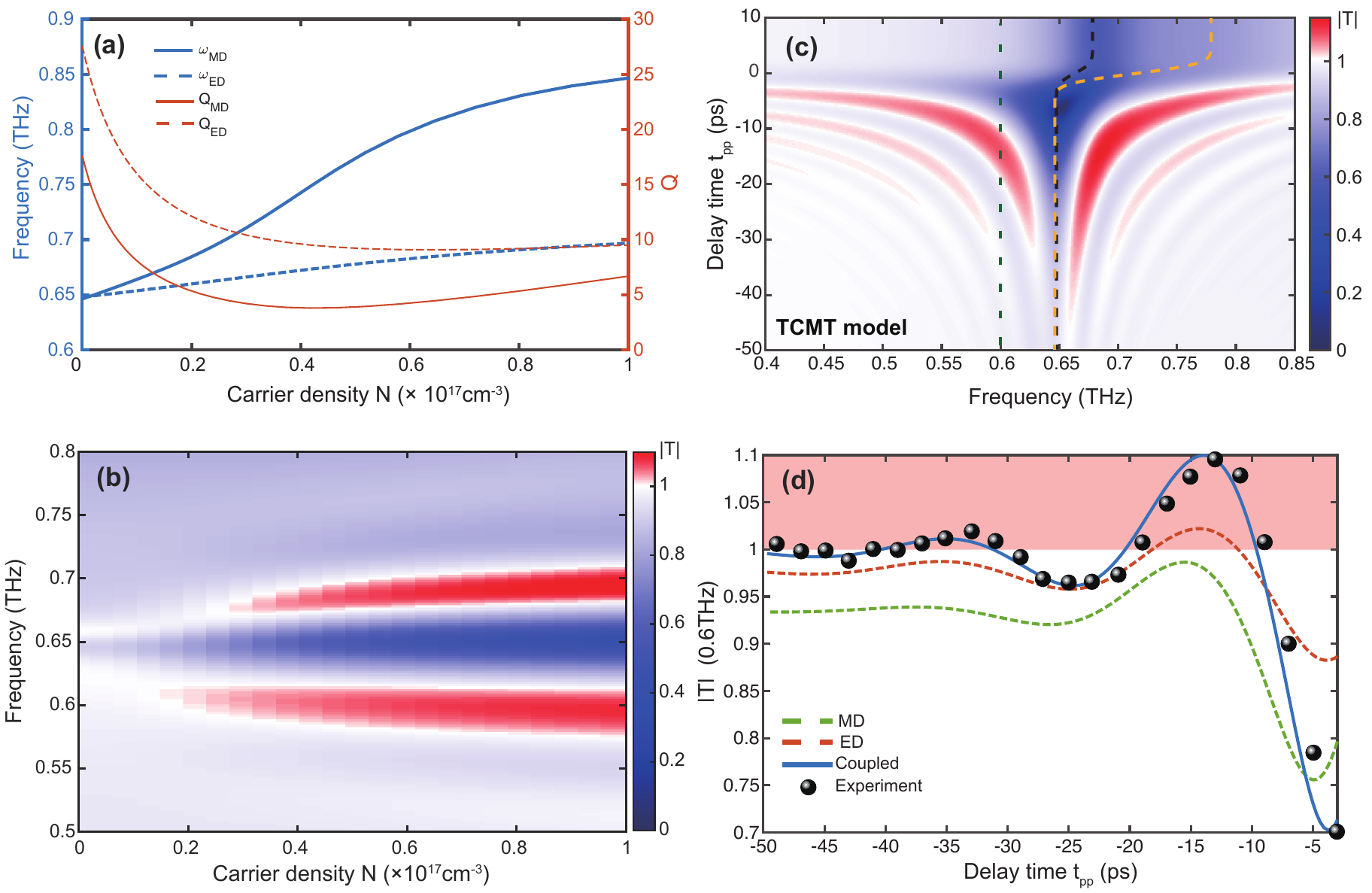}}
\caption{Theoretical analysis of the time-boundary-engineered Huygens’ metasurface. (a) Dependence of the resonant frequency and \textit{Q}-factor of the Huygens’ metasurface on carrier density in the photo-active layer. Blue lines show evolution of ED and MD resonant frequencies, and red lines the associated \textit{Q}-factors. (b) Simulated transmission spectra as a function of carrier density in the top layer, at the pump-probe delay of -10 ps. (c) Time-dependent transmission spectra obtained from TCMT model using the parameters from simulated eigenvalue of coupled ED and MD modes. The black and orange dashed traces indicate the temporal evolution of the ED and MD resonant frequencies. Green dashed curve shows frequency for maximum gain in the experiment. (d) Temporal evolution of the transmission amplitude at 0.6 THz extracted from experimental results (black spheres, 20 $\mu$J/cm$^2$ excitation fluence), TCMT analysis involving a single MD (dashed green line) or single ED (dashed red line) mode or coupled modes (solid blue line).}
\label{Fig4}
\end{figure}

\end{document}